\documentclass[prl,preprint,superscriptaddress,altaffillsymbol,raggedfooter]{revtex4}
\usepackage{graphicx}
\usepackage{dcolumn}
\usepackage{amssymb}
\usepackage{bm}
\usepackage{pifont}
\usepackage{epsfig}
\usepackage{psfrag}
\usepackage{epstopdf}
\usepackage{wasysym}
\usepackage[usenames]{color}

\begin{document}

\title{Growing Surface Tension of Amorphous-Amorphous Interfaces on Approaching the Colloidal Glass Transition}
\author{Divya Ganapathi{$^*$}}
\affiliation{Department of Physics, Indian Institute of Science, Bangalore - 560012, INDIA}
\author{Hima K Nagamanasa}
\affiliation{Chemistry and Physics of Materials Unit, Jawaharlal Nehru Centre for Advanced Scientific Research, Jakkur, Bangalore - 560064, INDIA}
\altaffiliation{Present Address: IBS Center for Soft and Living Matter, UNIST, Ulsan 689-798, South Korea}
\author{A K Sood}
\affiliation{Department of Physics, Indian Institute of Science, Bangalore - 560012, INDIA}
\affiliation{International Centre for Materials Science, Jawaharlal Nehru Centre for Advanced Scientific Research, Jakkur, Bangalore - 560064, INDIA}
\author{Rajesh Ganapathy{$^+$}}
\affiliation{International Centre for Materials Science, Jawaharlal Nehru Centre for Advanced Scientific Research, Jakkur, Bangalore - 560064, INDIA}
\affiliation{Sheikh Saqr Laboratory, Jawaharlal Nehru Centre for Advanced Scientific Research, Jakkur, Bangalore - 560064, INDIA}
\date{\today}

\begin{abstract}

There is mounting evidence indicating that relaxation dynamics in liquids approaching their glass transition not only becomes increasingly cooperative \cite{BB_RevModPhys,Biroli_NatPhys} but the relaxing regions also become more compact in shape \cite{Kob_NatPhys,Berthier2_PRE,Szamel_JCP,Hima_NatPhys,chandan_PRL}. While the surface tension of the interface separating neighboring relaxing regions is thought to play a crucial role in deciding both their size and morphology \cite{RFOT_1,Lubchenko_Rev,Wolynes_NatPhys}, owing to the amorphous nature of these regions, even identifying these interfaces has not been possible in bulk liquids. Here, by devising a scheme to identify self-induced disorder sites in bulk colloidal liquids, we directly quantified the dynamics of interfaces delineating regions of high and low configurational overlap. This procedure also helped unveil a non-monotonicity in dynamical correlations that has never been observed in bulk supercooled liquids. Using the capillary fluctuation method \cite{Henk_Sci,Weeks_PNAS}, we measured the surface tension of amorphous-amorphous interfaces with supercooling and find that it increases rapidly across the mode-coupling area fraction. Remarkably, a similar growth in the surface tension is also seen in the presence of a pinned amorphous wall. Our observations help prune theories of glass formation and opens up new research avenues aimed at tuning the properties of amorphous-amorphous interfaces, and hence the glass itself, in a manner analogous to grain boundary engineering in polycrystals \cite{GB_book}.

\end{abstract}
\maketitle

The seminal observation that the dynamical equations in certain spin glass models shared profound similarities with mode-coupling theory (MCT) of real liquids \cite{Gotze_MCT} led Kirkpatrick, Thirumalai and Wolynes to formulate the random first-order theory (RFOT) of glasses \cite{RFOT_1,Lubchenko_Rev}. While in MCT the structural relaxation time $\tau_\alpha$ diverges at $T_c$ (the mode-coupling temperature), within RFOT, $T_c$ is a dynamic crossover below which the liquid fragments into a patchwork of \textit{distinct} amorphous mosaics and relaxation is via activated mosaic rearrangements. The competition between the configurational entropy gain and the interfacial energy cost following a rearrangement sets the size of these mosaics, which in turn governs $\tau_\alpha$. With further supercooling, the mosaic size is expected to grow and eventually diverge at a bona fide thermodynamic transition to an ideal glass at $T_K$ \cite{BB_RevModPhys, Lubchenko_Rev,Gokhale_AdvPhy}. Although a nonzero surface tension, $\Upsilon$, is essential for the stability of these mosaics \cite{Cammarota_JStat}, even identifying these interfaces, let alone quantify the evolution of $\Upsilon$ across $T_c$, has not been possible in bulk liquids. Instead, motivated by measurements of the mosaic size, also called the point-to-set length $\xi_{PTS}$, in particle pinning based methods \cite{Biroli_NatPhys}, theoretical and numerical studies have attempted to use this method to quantify $\Upsilon$ \cite{Cammarota_JStat, Franz_JStat} and also probe the statistics of interface fluctuations \cite{Biroli_PRX}. While $\Upsilon$ was estimated to grow monotonically across $T_c$, the study focused on inherent structures and moreover measuring $\Upsilon$ also involved swapping particles within a cavity keeping the boundary ones frozen and hence has no experimental analogue \cite{Cammarota_JStat}. Simulations that probed the dynamics of a liquid near a pinned amorphous wall have also uncovered direct evidence for the change in relaxation  mechanism across $T_c$ in terms of a maximum in the dynamic correlation length, $\xi_d$ \cite{Kob_NatPhys}. Subsequent colloid experiments that mimicked the simulation protocol not only corroborated these findings but also showed that the maximum in $\xi_d$ coincides with the change in shape of most-mobile particle clusters from string-like to compact \cite{Hima_NatPhys}. This observation was at odds with the purely kinetic dynamical facilitation theory of glasses \cite{Gokhale_AdvPhy,Keys_PRX}.

Even while the artificial introduction of disorder by pinning particles seems to be a prerequisite for testing predictions from RFOT \cite{Biroli_NatPhys,Cammarota_JStat,Franz_JStat,Biroli_PRX,Kob_NatPhys,Hima_NatPhys,Berthier_PRE2012}, whether these findings readily carry over to structural glasses, where disorder is \textit{self-pinned} \cite{BB_RevModPhys}, remains unanswered. At present, even in numerical studies on bulk supercooled liquids, evidence for a change in relaxation dynamics across $T_c$ is rather indirect \cite{Berthier2_PRE,Szamel_JCP} leading to suggestions that the non-monotonicity in $\xi_d$ may be unique to the pinned wall geometry \cite{Flenner_NatPhys,Mei_PRE}. These concerns notwithstanding, simulations find that a pinned wall can subtly influence particle dynamics by exerting entropic forces that depend on the nature of the inter-particle potential \cite{Benjamin_PRE}. Similar problems persist even when the particles are randomly pinned. Although increasing the fraction of the pins results in a substantial growth in $\tau_\alpha$, the peak in the dynamic susceptibility, $\chi^{*}_4$, related to the size of dynamic heterogeneities, remains nearly constant unlike in bulk liquids where $\chi^{*}_4$ steadily grows with supercooling \cite{Jack_PRE}; implying that the nature of slowing down in the pinned liquid may be quite different from the bulk. Dynamics aside, there is no consensus on whether $\xi_{PTS}$ is even order agnostic and tracks structural correlations that are different from those obtained from simple pair-correlations \cite{Tanaka_PNAS,Sho_PRE}. Developing strategies to help resolve these controversies is a much needed step towards solving the glass transition puzzle. 

\begin{figure}[htbp]
\includegraphics[width=0.6\textwidth]{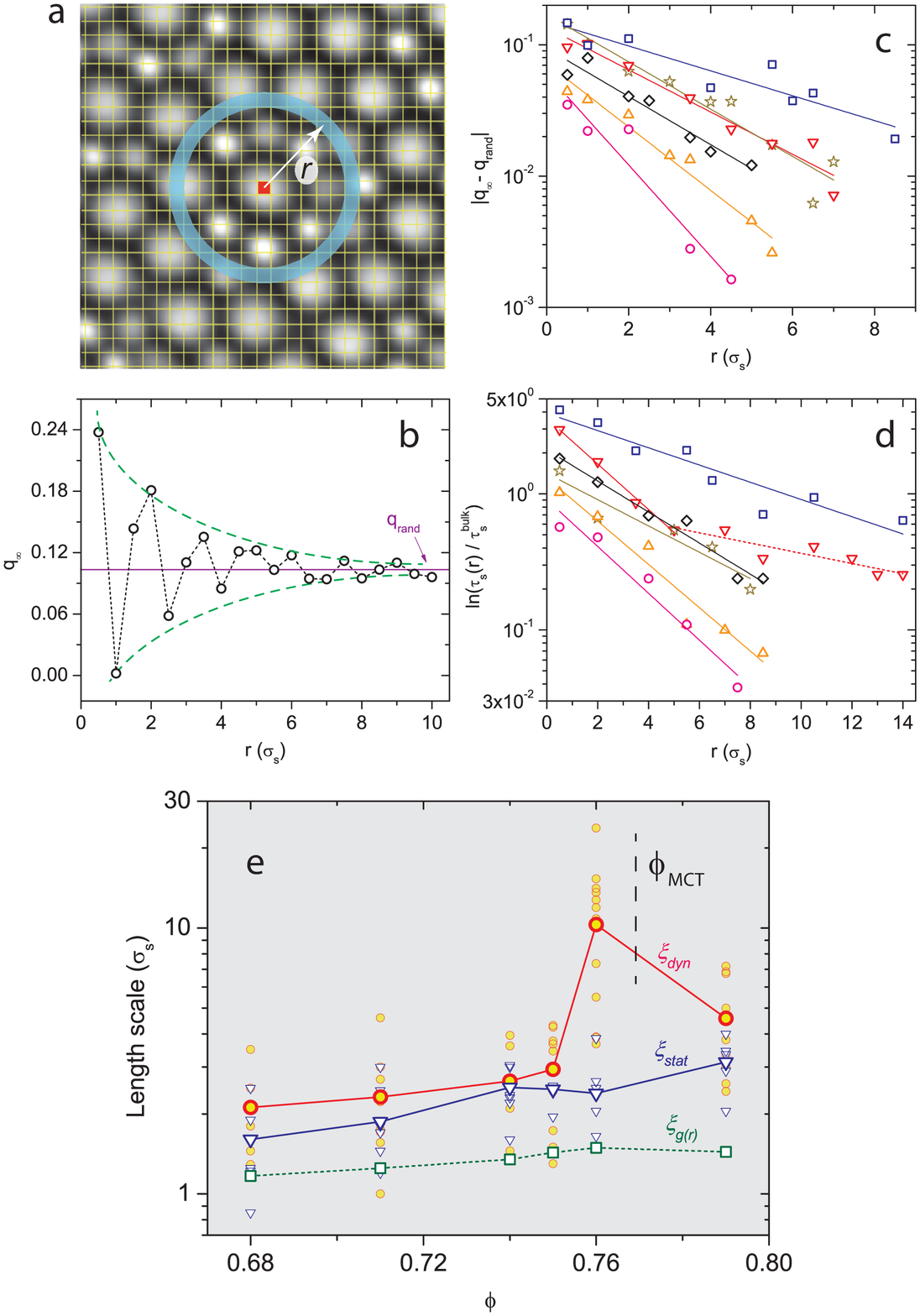}
\caption{\textbf{Growing static and non-monotonic dynamic correlations around self-induced pins.} (a) The background image represents a small portion of the field of view of a colloidal supercooled liquid. The red box in the center of the image has $q_c(\tau_\alpha)>0.9$ and represents a self-induced pin. $\langle q_c(t)\rangle_r$ is obtained by radially averaging boxes at a distance $r$ (blue ring) from the center. (b) shows the evolution of $q_\infty$ around a self-induced pin. The purple line corresponds to the bulk value $q_{rand}$. (c) and (d) correspond to  $|q_{\infty}(r) - q_{rand}|$ and $\ln({\tau_s(r)\slash\tau_s^{bulk}})$ versus $r$, respectively. $\phi = 0.68$ (circles), $\phi = 0.71$ (upright triangles), $\phi = 0.74$ (diamonds), $\phi = 0.75$ (diamonds), $\phi = 0.76$ (inverted triangles), $\phi = 0.79$ (squares). (e) The small circles and triangles represent $\xi_d$ and $\xi_{stat}$, respectively, for many independent pins for a given $\phi$ and the larger symbols represent their averages. The squares correspond to the two-point correlation length.}
\label{Figure1}
\end{figure}

Here we devised a method to identify self-induced pins and probe their influence on local structure and dynamics in bulk supercooled colloidal liquids. This conceptual advance also enabled us to measure the surface tension of amorphous-amorphous interfaces. Single-particle dynamics at different colloidal particle area fractions, $\phi$, were studied using optical video microscopy (see Methods). We identified self-induced pins by exploiting the fact that $\tau_\alpha$ is determined by the slowest relaxing regions in the supercooled liquid and hence regions that harbor these pins should also be configurationally similar over at least $\tau_\alpha$. The configurational overlap $q_c(t)$ which measures the extent of this similarity over time is also thought to be the order parameter within RFOT \cite{Lubchenko_Rev,BB_RevModPhys}. We measured $q_c(t)$ by first coarse-graining the field of view into boxes of size 0.5$\sigma_s$ (Fig. \ref{Figure1}a). The box size is optimally chosen to minimize overlap fluctuations due to cage rattling as well as avoid multi-particle occupancy in a given box \cite{Kob_NatPhys,Hima_NatPhys}. Nevertheless, particles located near the box edges can still hop to neighboring boxes and this results in spurious overlap fluctuations. We accounted for these by developing the fuzzy-grid method which involves displacing the coarse-graining grid by the cage size $0.1\sigma_s$ in various directions and then averaging $q_c(t)$ obtained from each of the realizations (see Supplementary text). For each box and for $t=\tau_\alpha$, we computed $q_c(t) = \frac{\langle n_i(t)n_i(0)\rangle_t}{\langle n_i(0)\rangle_t}$, where $i$ is the box index, $n_{i}(t) = 1$ if the box contains a particle at time $t$ and $n_{i}(t) = 0$ otherwise. Unlike the case of quenched disorder, self-induced pins do not persist indefinitely since the liquid eventually relaxes and hence the time averaging denoted by $\langle\rangle_t$ is performed over $1\tau_\alpha$. Boxes with a $q_c(\tau_\alpha) > 0.9$ were identified as self-induced pins (red box in Fig. \ref{Figure1}a and Supplementary Fig. S2) and we restricted our attention to those that persisted over many consecutive $\tau_\alpha$'s. For $\phi = 0.79 >\phi_{MCT}$, however, owing to experimental difficulties with sample equilibration, $t = 7t^*$. 

Having identified the pins, we probed their influence on local static order and dynamics by adapting the procedure originally developed for the amorphous wall geometry \cite{Kob_NatPhys}. We computed the radially averaged configurational overlap $q_c(t,r) = \langle q_c(t)\rangle_r$ and the self-overlap $q_s(t,r) = \langle{\frac{\langle n_{i}^{s}(t)n_{i}^{s}(0)\rangle_t} {\langle n_{i}^{s}(0)\rangle_t}\rangle_r}$ for all boxes at a distance $r$ from a given pin. Here, $n_{i}^{s}(t) = 1$ if the box is occupied by the \textit{same} particle at time $t$ and $n_{i}^{s} = 0$ otherwise. By construction $q_c(t,r)$ is insensitive to particle swaps, while $q_s(t,r)$ is. Supplementary Fig. S3a and b show the time evolution of $q_c(t,r)$ and $q_s(t,r)$ for $\phi = 0.75$ at different $r$'s, respectively. The long time limit of $q_c(t\to \infty,r) = q_{\infty}$ saturates to the bulk value $q_{rand}$ for large $r$'s, while closer to the pin $q_\infty$ is larger (Fig. \ref{Figure1}b) indicating that density fluctuations are frozen. $q_{rand}$ measures the probability that a box is occupied. The oscillations in $q_\infty$ are due to local liquid-like ordering around the pin. Strikingly, analogous to observations on externally-pinned liquids \cite{Kob_NatPhys,Hima_NatPhys,Berthier_PRE2012}, the excess contribution to the configurational overlap over the bulk, $|q_\infty(r) - q_{rand}|$, decayed exponentially with $r$ for all $\phi$'s. We estimated a static correlation length $\xi_{stat}$ from the relation $|q_{\infty}(r) - q_{rand}| = B\text{exp}(-r/\xi_{stat})$ \cite{Kob_NatPhys,Hima_NatPhys}. Unlike $q_c(t,r)$, $q_s(t,r)$ decays to zero in the long time limit, when particles have undergone displacements larger than the box size (Supplementary Fig. S3b). We defined the relaxation time $\tau_s(r)$ as the time taken for $q_s(t,r)$ to decay to $0.3$. As expected, $\tau_s(r)$ close to the pin is larger than its bulk value $\tau_s^{bulk}$. In Fig. \ref{Figure1}d, we plot $\ln({\tau_s(r)\slash\tau_s^{bulk}})$ versus $r$ for different $\phi$'s. For all $\phi$'s except $\phi = 0.76$ (inverted red triangles in Fig. \ref{Figure1}d), $\ln({\tau_s(r)\slash\tau_s^{bulk}})$ showed an exponential decay. For $\phi = 0.76$, which is close to the mode-coupling area fraction ($\phi_{MCT} = 0.77$, see Supplementary Fig. S4), however, we clearly see two slopes. Such a departure from an exponential decay has been attributed to the presence of multiple relaxation processes near the MCT transition \cite{Wolynes_NatPhys,Kob_NatPhys,Hima_NatPhys}. We extracted a dynamic length scale $\xi_d$ from $\ln({\tau_s(r)\slash\tau_s^{bulk}}) = B_s\text{exp}(-r/\xi_{d})$ \cite{Kob_NatPhys}.

We performed the equivalent of disorder averaging by repeating the above analysis for many well-separated pins within our field of view. The small hollow triangles and circles in Fig. \ref{Figure1}e correspond to $\xi_{stat}$ and $\xi_d$ evaluated for each of these pins and the larger symbols correspond to their averages for each $\phi$. We also show the two-point correlation length $\xi_{g(r)}$ (hollow green squares) obtained by fitting an envelope to the decay of the pair correlation function $g(r)$ for comparison. While $\xi_{stat}$ clearly grows faster than $\xi_{g(r)}$, the more striking feature is the presence of a maximum in $\xi_d$ near $\phi_{MCT}$. This is the first observation of a non-monotonic evolution of a dynamic length scale in a bulk supercooled liquid and more importantly, is also consistent with the change in shape of dynamical heterogeneities observed earlier \cite{Hima_NatPhys}. Interestingly, we also observe that the spread in $\xi_d$ values is maximal near $\phi_{MCT}$ (Supplementary Fig. S5) and this may be another signature of being in the vicinity of the dynamical crossover predicted by RFOT.

Inspired by the similarities in the $\phi$-dependence of $\xi_{stat}$ and $\xi_d$ between artificially introduced and self-induced pinning, we explored if methods for identifying amorphous-amorphous interfaces in the former can be extended to the latter \cite{Cammarota_JStat, Franz_JStat,Biroli_PRX}. Since experimental studies of such interfaces is lacking even for pinned liquids, we first analysed data on colloidal liquids in the presence of an optically pinned amorphous wall \cite{Hima_NatPhys} (see supplementary Fig. S6). Figure \ref{Figure2}a shows a snapshot of $q_c(t=\tau_\alpha)$ calculated for a coarse-graining grid size of $1\sigma_s$ for $\phi = 0.74$. The image corresponds to the portion of the field of view that contains the wall located at $z\leq0$ (dashed line). In line with expectations, $q_c(\tau_\alpha)$ for regions close to the wall ($z>0$) is larger as compared to regions farther away from it (represented by green and yellow boxes, respectively). Since the grid size is smaller than the average particle size, boxes in the vicinity of those with a high $q_c(\tau_\alpha)$ necessarily have low $q_c(\tau_\alpha)$ values and are thus equivalent. Thus, boxes with $q_c(\tau_\alpha)\leq 0.15 \sim q_{rand}$ are also colored green in the image. Next, starting from each box at $z=0$ we scanned along $z$ and located the box where the overlap dropped to $0.15\leq q_c(\tau_\alpha)\leq 0.67$. The line joining these boxes delineates regions of high and low configurational overlap \cite{Franz_JStat,Biroli_PRX} and we defined this to be the instantaneous interface profile, $h(z,t)$, for a given $\tau_\alpha$ (black line in Fig. \ref{Figure2}a). We followed this procedure for each $\tau_\alpha$ and quantified the dynamics of $h(z,t)$. The time-averaged interface profile, $\langle h(z,t)\rangle_t$, as expected, was found to be parallel to the wall at $z=0$ (white line in Fig. \ref{Figure2}a). For each $\phi$, interface fluctuations were probed over their corresponding $\tau_\alpha$ and a direct comparison of their dynamics is thus possible. Supplementary video S1 and S2 show interface fluctuations for the pinned wall geometry for $\phi = 0.75$ and $\phi = 0.79$, respectively.

We made modifications to the above procedure for identifying interfaces around self-induced pins. Starting from the pin (represented as solid white circle in Fig. \ref{Figure2}b) we scanned along $x$ and $z$ directions for boxes where the overlap dropped to $0.15\leq q_c(\tau_\alpha)\leq 0.67$ (see Supplementary text and Fig. S7). A line through these boxes represents the instantaneous interface profile. Figure \ref{Figure2}c shows interfaces obtained from this procedure for many well-separated pins within our field of view for $\phi = 0.79$. At high $\phi$'s especially, regions with a high configurational overlap typically contain more than one self-induced pin (see Supplementary Fig. S2). We have checked that the interface profile obtained starting from any of these pins defines nearly the same high overlap region (white and pink interface profiles around pins labeled 1 and 2, respectively in Fig. \ref{Figure2}c). We then carried out the fuzzy-grid averaging procedure to smoothen out interface fluctuations for both the pinned wall and the self-induced pins. The resulting interface profile was interpolated (black line through red symbols in Fig. \ref{Figure2}d) before further analysis. Supplementary video S3 and S4 show interface fluctuations around self-induced pins for $\phi = 0.75$ and $\phi = 0.79$, respectively.

The surface tension of an interface is inversely related to its roughness which is best captured by the interface width $w = \sqrt{(\Delta h)^2} = \sqrt{(h(z,t) - \langle h(z,t)\rangle_t)^2}$. Since the amplitude of interface fluctuations is extensive in the system size and diverges in the thermodynamic limit, only interfaces of the same length were considered (see Supplementary methods). Figure \ref{Figure3}a and Supplementary Fig. S8 shows the evolution in the distribution of height fluctuations, $P(\Delta h)$ with $\phi$, for the self-induced pin and the pinned wall, respectively. $P(\Delta h)$ gets narrower with $\phi$ which already signals a growth in the $\Upsilon$ of the high $q_c$ regions. While $P(\Delta h)$ is fully captured by Gaussian fits at low $\phi$'s, deviations were observed for $\phi > 0.75$ and $w$ was extracted by fitting only the central region where $P(\Delta h)$ dropped by about a decade. In Fig \ref{Figure3}b, we show $w$ versus $\phi$ for the self-induced pin (solid symbols) and the pinned wall (hollow symbols), respectively. Strikingly, for both these cases we observed that $w$ appears to taper of beyond $\phi_{MCT}$.

Next, we attempted to measure $\Upsilon$ directly using the capillary fluctuation method (CFM). CFM was originally developed to quantify dynamics of \textit{flat} interfaces separating phases with a well-defined order parameter \cite{Henk_Sci,Weeks_PNAS} and it is not immediately apparent if this approach can be extended to configurational overlap fields and that too for interfaces that are curved like those seen around self-induced pins. At finite temperature, interfaces undergo broadening due to thermal fluctuations and the final equilibrium profile is a trade off between the surface energy term which prefers a flat interface and the thermal energy $k_BT$. In CFM, $h(z,t)$ is decomposed into normal modes and the amplitude of each mode decays as $\langle| A(k)|^2\rangle = {k_BT\over{L\Upsilon k^2}}$ in accordance with the equipartition theorem. Here, $k$ is the wavevector and $L$ is the length of the interface. Figure \ref{Figure3}c and Supplementary Fig. S9 show $L\langle| A(k)|^2\rangle k^2$ versus $k$ as a function of $\phi$ for a representative self-induced pin and the pinned wall, respectively. Remarkably, we observe that $L\langle| A(k)|^2\rangle k^2$ is constant for almost a decade in $k$ which not only validates the applicability of CFM for the present system but also allowed us to quantify $\Upsilon$ for the first time in experiments. Figure \ref{Figure3}d shows $\Upsilon$ versus $\phi$ for many well-separated self-induced pins (small gray circles) and the pinned wall (hollow squares). The filled black circles correspond to the average $\Upsilon$ for each $\phi$. Remarkably, in both cases, we observe that with supercooling $\Upsilon$ grows rather rapidly in the vicinity of $\phi_{MCT}$. We ensured that our findings are not very sensitive to the box size as well as the precise values of the overlap limits chosen (see supplementary text and Fig. S10). Such a growth in $\Upsilon$ should immediately result in CRRs shapes which become increasingly compact across $\phi_{MCT}$ \cite{Kob_NatPhys,Hima_NatPhys} and is consistent with theoretical predictions \cite{Wolynes_NatPhys}. Further, in the mean-field limit, $\Upsilon$ is expected to be zero beyond the spinodal singularity, loosely identified with $T_{c}$, which is also the upper bound for RFOT-like mechanisms to be relevant. In finite dimensions, however, numerical studies find that the spinodal singularity is not sharply defined and $\Upsilon$ vanishes smoothly for $T>T_{c}$ \cite{Cammarota_JStat}. This is once again consistent with our observations of a finite $\Upsilon$ for $\phi<\phi_{MCT}$. 

We finally focused on the dynamics of most-mobile particles at amorphous-amorphous interfaces. In Fig \ref{Figure3}e, we show the top $1\%$ of particles that were labelled most-mobile over a $t^*$ interval \cite{Weeks_Science} that fell within the time window ($7t^*$) over which the underlying configurational overlap was calculated for $\phi = 0.79$. The different colors represent different particles. While it is not surprising that these particles are predominatly found in the low $q_c$ regions, we observe that for most particles the trajectories are elongated along the interface length. This anisotropic weakening of the cage should result in string-like CRRs oriented along the interface and is strikingly similar to the dynamics of particles at crystal grain boundaries \cite{Hima_PNAS}.

Our observations of a rapid increase in $\Upsilon$ of amorphous-amorphous interfaces and a decrease in $\xi_d$ on supercooling across $\phi_{MCT}$ can be currently reconciled only with the framework of RFOT. In the ongoing quest for identifying the relevant length scale(s) that best capture the growth in $\tau_\alpha$ in the $T<T_{c}$ ($\phi>\phi_{MCT}$) regime \cite{Flenner_NatPhys}, determining whether $\xi_d$ is eventually slaved to $\xi_{stat}$ is crucial. While this is presently beyond the scope of particle-resolved experiments, numerical studies that exploit the recently developed swap Montecarlo technique may be the way forward \cite{Berthier_Swap}. Finally, given that the involved procedure of artificially pinning particles is not essential to distinguish between competing theories of glass formation opens the door to extending our method to supercooled liquids made of particles with complex shapes and internal degrees of freedom.  

\section*{Materials and Methods}
Our experimental system comprised of a binary mixture of colloidal polystyrene particles with sizes $\sigma_s$=1.05$\mu$m and $\sigma_L$=1.4$\mu$m. The number ratio $N_L/N_S =$ 1.23 was sufficient to prevent crystallization and was held nearly constant for all $\phi$'s studied. The colloidal suspensions were loaded in a wedge-shaped cell that was left standing for a suitable time duration to yield the desired particle area fraction $\phi$. Here, $\phi$ plays the role of an inverse temperature. The systems were equilibrated for a typical time duration of 8-10h before the experiments (several times $\tau_\alpha$ for all $\phi s<0.79$).  Samples were imaged using a Leica DMI 6000B optical microscope with a 100X objective (Plan apochromat, NA 1.4, oil immersion) and images were captured at frame rates ranging from 3.3 fps to 5 fps for 1-1.5 hours depending on the values of $\phi$. The typical field of view captured in our experiment is of the size 51$\sigma_s \times 39\sigma_s$. The analysis reported here were carried out on experiments that were performed immediately after the holographic optical tweezers that was used to pin an amorphous wall of particles was turned off \cite{Hima_NatPhys} (See supplementary text). Thus, a direct comparison of the present results with the earlier study is thus justified. The particle trajectories were obtained from standard MATLAB algorithms \cite{crocker1996methods}. Subsequent analysis was performed using codes developed in-house. The typical drift observed in the experimental data in the presence of a pinned amorphous wall was of the order of 0.7$\sigma_s$ in the x-direction and 0.15$\sigma_s$ in the z-direction over the entire duration of the experiment. Since sample drift corrections are not possible in the presence of the wall, we choose the longest time window where the sample drift was less than 0.1$\sigma_s$ for further analysis.

\section*{Acknowledgements}
We thank Walter Kob for illuminating discussions and Francesco Sciortino for encouraging comments. We thank Shreyas Gokhale for critical feedback on our manuscript. A.K.S thanks J C Bose Fellowship of the Department of Science and Technology (DST), India for support. R.G. thanks the ICMS and SSL, JNCASR for financial support and DST for a SwarnaJayanti Fellowship grant.

\section*{Author Contributions}
D.G., A.K.S. and R.G. conceived research. H.K.N. performed experiments. D.G. and R.G. developed analysis procedures. D.G. performed analysis. R.G. wrote the paper with inputs from A.K.S.

$^*$,$^+$ Corresponding author

\newpage

\newpage
\begin{figure}[htbp]
\includegraphics[width=1\textwidth]{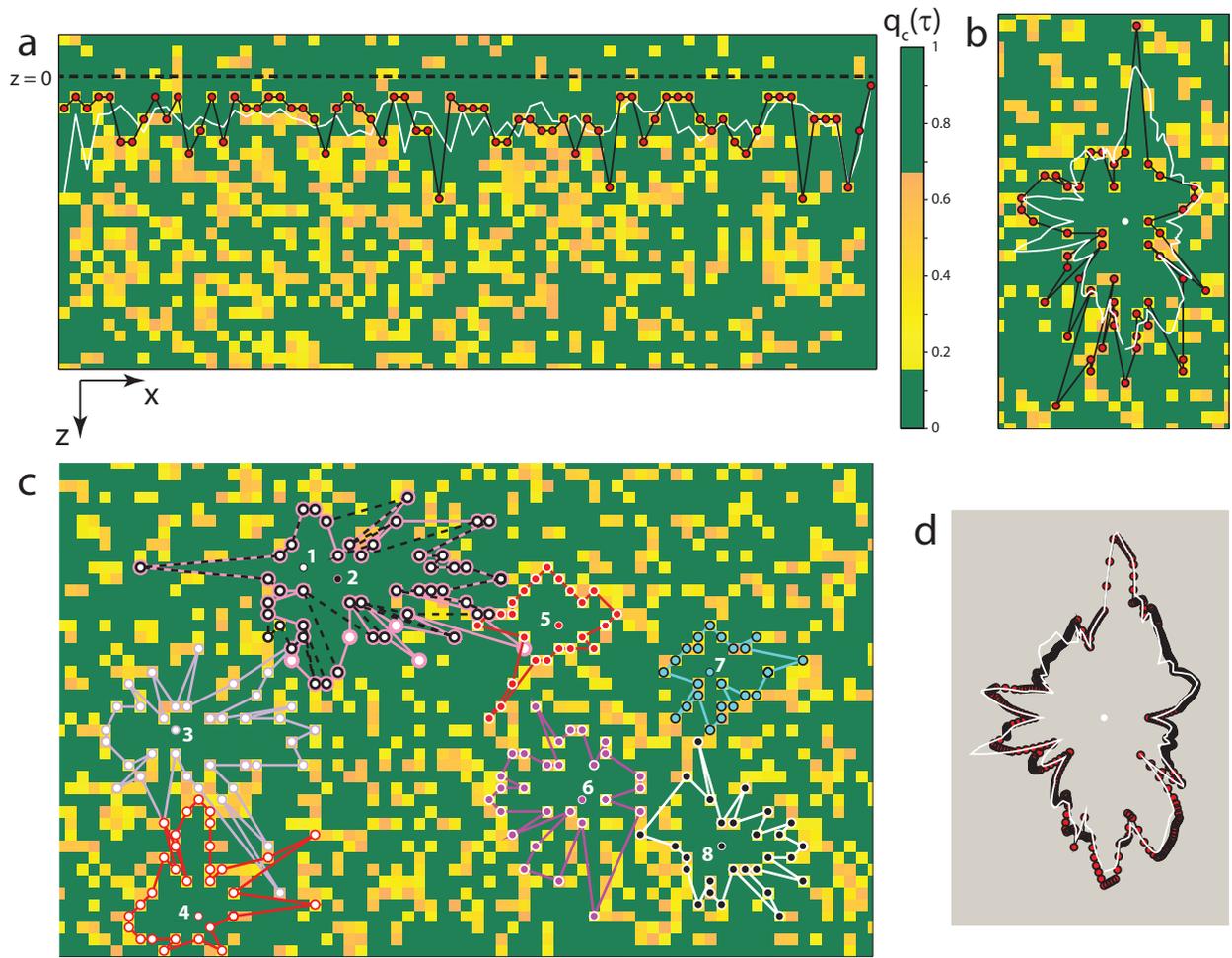}
\caption{\textbf{Identifying amorphous-amorphous interfaces.} (a) The back ground image represents the $q_c(\tau_\alpha)$ for a portion of the field of view containing the amorphous wall located at $z\leq0$ (dashed line) for $\phi = 0.75$. The color bar represents $q_c(\tau_\alpha)$ values. The black line through the red symbols corresponds to the instantaneous interface profile $h(z,t)$ and the white line is the time-averaged profile $\langle h(z,t)\rangle_t$. (b) Interface profile around self-induced pin represented by the white circle. The background color scheme and the lines have the same meaning as in (a). (c) Instantaneous interface profiles around distinct self-induced pins (represented by the numbered circles). Pins 1 and 2 lie within the same mosaic and yield nearly the same interface. (d) Instantaneous interface profile post fuzzy-grid averaging. In (b)-(d) $\phi = 0.79$.}
\label{Figure2}
\end{figure}

\newpage
\begin{figure}[htbp]
\includegraphics[width=0.75\textwidth]{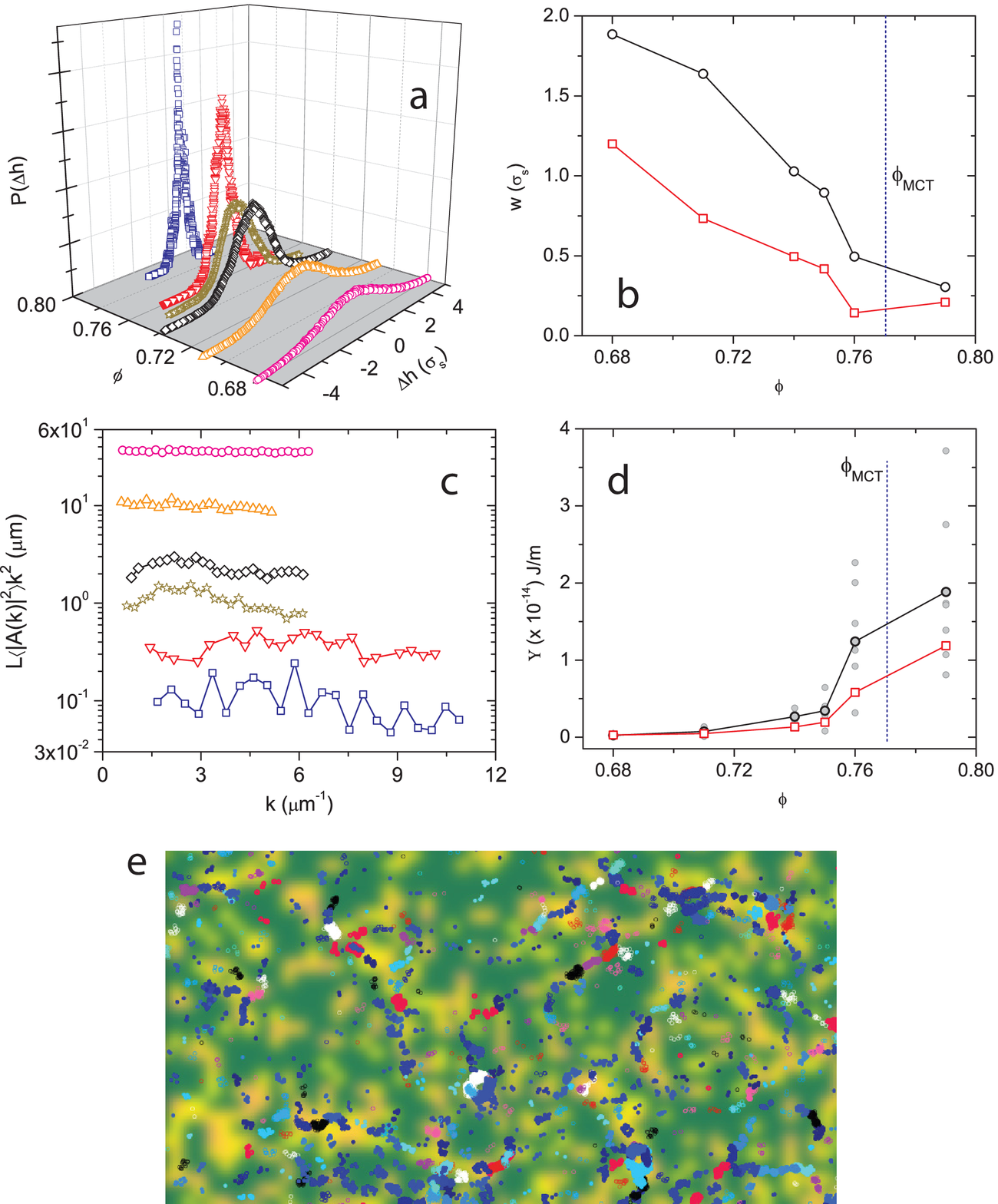}
\caption{\textbf{Surface tension of amorphous-amorphous interfaces.} (a) Distribution of height fluctuations as a function of $\phi$. $\phi = 0.68$ (circles), $\phi = 0.71$ (upright triangles), $\phi = 0.74$ (diamonds), $\phi = 0.75$ (diamonds), $\phi = 0.76$ (inverted triangles), $\phi = 0.79$ (squares). (b) Interface width $w$ versus $\phi$ for amorphous wall (hollow squares) and for a representative self-induced pin (hollow circles). (c) $L\langle| A(k)|^2\rangle k^2$ versus $k$ for a representative self-induced pin. The symbol shape correspond to same $\phi$'s as in (a). (d) the small hollow circles represent $\Upsilon$ for distinct self-induced pins and the large circles represents their average for each $\phi$. $\Upsilon$ versus $\phi$ for amorphous wall (hollow squares). (e) The background image represents $q_c(t = 7t^*)$ for $\phi = 0.79$. The trajectories of the top 1\% most-mobile particles are shown by the colored symbols.}
\label{Figure3}
\end{figure}

\begin{references}

\bibitem{BB_RevModPhys} Berthier, L., and Biroli, G., Theoretical perspective on the glass transition and amorphous materials. \textit{Reviews of Modern Physics} \textbf{83}, 587 (2011).

\bibitem{Biroli_NatPhys} Biroli, G., et al. Thermodynamic signature of growing amorphous order in glass-forming liquids. \textit{Nature Physics} \textbf{4} 771-775 (2008).

\bibitem{Kob_NatPhys} Kob, W., Roldan-Vargas, S., and Berthier, L., Non-monotonic temperature evolution of dynamic correlations in glass-forming liquids. \textit{Nature Physics} \textbf{8}, 164-167 (2012).

\bibitem{Berthier2_PRE} Berthier, L., et al., Finite-size effects in the dynamics of glass-forming liquids. \textit{Physical Review E} \textbf{86}, 031502 (2012).

\bibitem{Szamel_JCP} Flenner, E., and Szamel, G., Dynamic heterogeneities above and below the mode-coupling temperature: Evidence of a dynamic crossover. \textit{Journal of chemical physics} \textbf{138}, 12A523 (2013).

\bibitem{Hima_NatPhys} Nagamanasa, K. H., Gokhale, S., Sood, A. K., and Ganapathy, R., Direct measurements of growing amorphous order and non-monotonic dynamic correlations in a colloidal glass-former. \textit{Nature Physics} \textbf{11}, 403-408 (2015).

\bibitem{chandan_PRL} Mishra, C. K., and Ganapathy, R., Shape of dynamical heterogeneities and fractional Stokes-Einstein and Stokes-Einstein-Debye relations in quasi-two-dimensional suspensions of colloidal ellipsoids. \textit{Physical Review Letters}   
\textbf{114}, 198302 (2015).

\bibitem{RFOT_1} Kirkpatrick, T. R., Thirumalai, D., and Wolynes, P. G., Scaling concepts for the dynamics of viscous liquids near an ideal glassy state. \textit{Physical Review A} \textbf{40}, 1045 (1989).

\bibitem{Lubchenko_Rev} Lubchenko, V., and Wolynes, P. G., Theory of structural glasses and supercooled liquids. \textit{Annu. Rev. Phys. Chem.} \textbf{58}, 235-266 (2007).

\bibitem{Wolynes_NatPhys} Stevenson, J. D., Schmalian, J., and Wolynes, P. G., The shapes of cooperatively rearranging regions in glass-forming liquids. \textit{Nature Physics} \textbf{2}, 268-274 (2006).

\bibitem{Henk_Sci} Aarts, D. G. A. L., Schmidt, H., and Lekkerkerker, H. N. W.. Direct visual observation of thermal capillary waves. \textit{Science} \textbf{304}, 847-850 (2004).

\bibitem{Weeks_PNAS} Hernandez-Guzman, J., and Weeks, E. R., The equilibrium intrinsic crystal–liquid interface of colloids. \textit{Proceedings of the National Academy of Sciences} \textbf{106}, 15198-15202 (2009).

\bibitem{GB_book} Gottstein, G., and Shvindlerman, L.S., \textit{Grain boundary migration in metals: thermodynamics, kinetics, applications}. CRC press (2009).

\bibitem{Gotze_MCT} Bengtzelius, U., Gotze, w., and Sjolander, A., Dynamics of supercooled liquids and the glass transition. \textit{Journal of Physics C: solid state Physics} \textbf{17}, 5915 (1984).

\bibitem{Gokhale_AdvPhy} Gokhale, S., Sood, A. K., and Ganapathy, R., Deconstructing the glass transition through critical experiments on colloids. \textit{Advances in Physics} \textbf{65}, 363-453 (2016).

\bibitem{Cammarota_JStat} Cammarota, C., et al. Evidence for a spinodal limit of amorphous excitations in glassy systems. \textit{Journal of Statistical Mechanics: Theory and Experiment} \textbf{12}, L12002 (2009).

\bibitem{Franz_JStat} Zarinelli, E., and Franz, S., Surface tension in kac glass models. \textit{Journal of Statistical Mechanics: Theory and Experiment} \textbf{4}, P04008 (2010)

\bibitem{Biroli_PRX} Biroli, G., and Cammarota, C., Fluctuations and shape of cooperative rearranging regions in glass-forming liquids. \textit{Physical Review X} \textbf{7}, 011011 (2017).

\bibitem{Keys_PRX} Keys, A. S., et al. Excitations are localized and relaxation is hierarchical in glass-forming liquids. \textit{Physical Review X} \textbf{1}, 021013 (2011).

\bibitem{Berthier_PRE2012} Berthier, L., and Kob, W., Static point-to-set correlations in glass-forming liquids. \textit{Physical Review E} \textbf{85} 011102 (2012).

\bibitem{Flenner_NatPhys} Flenner, E., and Szamel, G., Characterizing dynamic length scales in glass-forming liquids. \textit{Nature Physics} \textbf{8}, 696-697 (2012).

\bibitem{Mei_PRE} Mei, B., et al. Nonmonotonic dynamic correlations in quasi-two-dimensional confined glass-forming liquids. \textit{Physical Review E} \textbf{95} 050601 (2017).

\bibitem{Benjamin_PRE} Benjamin, R., and Horbach, J., Excess free energy of supercooled liquids at disordered walls. \textit{Physical Review E} \textbf{90}, 060101 (2014).

\bibitem{Jack_PRE} Jack, R. L., and Fullerton, C. J., Dynamical correlations in a glass former with randomly pinned particles. \textit{Physical Review E} \textbf{88}, 042304 (2013).

\bibitem{Tanaka_PNAS} Russo, J., and Tanaka, H., Assessing the role of static length scales behind glassy dynamics in polydisperse hard disks. \textit{Proceedings of the National Academy of Sciences} \textbf{112}, 6920-6924 (2015).

\bibitem{Sho_PRE} Yaida, S., et al. Point-to-set lengths, local structure, and glassiness. \textit{Physical Review E} \textbf{94}, 032605 (2016).

\bibitem{crocker1996methods} Crocker, J. C. and Grier, D. G. Methods of digital video microscopy for colloidal studies. \textit{J. Colloid. Interface. Sci.} \textbf{179,} 298-310 (1996).

\bibitem{Weeks_Science} Weeks, E. R., et al. Three-dimensional direct imaging of structural relaxation near the colloidal glass transition. \textit{Science} \textbf{287}, 627-631 (2000).

\bibitem{Hima_PNAS} Nagamanasa, K. H., Gokhale, S., Ganapathy, R. and Sood, A. K., Confined glassy dynamics at grain boundaries in colloidal crystals. \textit{Proceedings of the National Academy of Sciences} \textbf{108}, 11323-11326 (2011).

\bibitem{Berthier_Swap} Berthier, L., et al.,  Breaking the glass ceiling: Configurational entropy measurements in extremely supercooled liquids. \textit{arXiv preprint arXiv:1704.08257}, (2017).

\end{references}
\end{document}